\documentclass[runningheads]{llncs}
\usepackage{graphicx}
\usepackage[T1]{fontenc} 
\usepackage[utf8]{inputenc} 
\usepackage{ifthen}
\usepackage{graphicx}
\usepackage{hyperref}
\usepackage{multirow}
\usepackage{array} 
\usepackage{arydshln} 
\usepackage[many]{tcolorbox} 
\usepackage{enumerate}
\usepackage{enumitem}

\usepackage{xspace}
\usepackage{xcolor}
\usepackage[normalem]{ulem} 			
\usepackage{amsmath,amsfonts}

\usepackage{caption} 
\usepackage{subcaption} 

\newcolumntype{L}[1]{>{\raggedright\let\newline\\\arraybackslash\hspace{0pt}}p{#1}}
\newcolumntype{C}[1]{>{\centering\let\newline\\\arraybackslash\hspace{0pt}}p{#1}}
\newcolumntype{R}[1]{>{\raggedleft\let\newline\\\arraybackslash\hspace{0pt}}p{#1}}

\newboolean{showcomments}
\setboolean{showcomments}{true}

\ifthenelse{\boolean{showcomments}}
{
	\newcommand{\nb}[3]{
		{\colorbox{#2}{\bfseries\sffamily\scriptsize\textcolor{white}{#1}}}
		{\textcolor{#2}{\sf\small$\blacktriangleright$\textit{#3}$\blacktriangleleft$}}}
	 
	\newcommand{\bnote}[2]{\fbox{\color{blue}\bfseries\sffamily\scriptsize#1}
    	{\color{blue}\sf\small$\blacktriangleright$\textit{#2}$\blacktriangleleft$}}
	\newcommand{\old}[1]{{\color{gray}\sout{#1}}} 
	\newcommand{\del}[1]{\old{#1}} 
	\newcommand{\ins}[1]{{\textcolor{blue}{\uline{#1}}}} 
	\newcommand{\ugh}[1]{{\textcolor{red}{\uwave{#1}}}} 
	\newcommand{\chg}[2]{{\textcolor{red}{\sout{#1}}}{\ra}\textcolor{blue}{\uline{#2}}} 
	 
	\newcommand{\fix}[1]{\bnote{FIX}{#1}}
}{
	\newcommand{\bnote}[2]{}
	\newcommand{\nb}[3]{}
	\newcommand{\old}[1]{}
	\newcommand{\del}[1]{}
	\newcommand{\ins}[1]{}
	\newcommand{\ugh}[1]{}
	\newcommand{\chg}[2]{}
	
	\newcommand{\fix}[1]{}
} 

\newcommand{\hide}[1]{}

\graphicspath{{img/}}

\newlist{RQ}{enumerate}{1}
\setlist[RQ]{label=RQ\arabic*:,leftmargin=3em}

\newlist{problems}{enumerate}{1}
\setlist[problems]{label=P\arabic*:,leftmargin=3em}

\newcommand{\commented}[1]{}

\newcommand{\etal}{\emph{et al.,}\xspace}

\usepackage{url}            
\makeatletter
\def\url@leostyle{%
  \@ifundefined{selectfont}{\def\UrlFont{\sf}}{\def\UrlFont{\small\sffamily}}}
\makeatother
\definecolor{main}{HTML}{828282}    
\definecolor{sub}{HTML}{E0E0E0}     

\tcbset{
    sharp corners,
    colback = white,
    before skip = 0.2cm,    
    after skip = 0.5cm      
}                           

\newtcolorbox{cbox}{
    enhanced, 
    boxrule = 0pt, 
    borderline = {0.75pt}{0pt}{main}, 
    borderline = {0.75pt}{2pt}{sub} 
}

\begin{document}
\title{Cormas: The Software for Participatory Modelling and its Application for Managing Natural Resources in Senegal\thanks{This work was sponsored by the CASSECS project: \url{https://www.cassecs.org/}.}}
%
%
\author{Oleksandr Zaitsev\inst{1}\orcidID{0000-0003-0267-2874} \and
François Vendel\inst{1,2} \and
Etienne Delay\inst{1,3}}
\authorrunning{O. Zaitsev et al.}
%
\institute{CIRAD, UMR SENS, MUSE, Université de Montpellier, France \\ \email{\{oleksandr.zaitsev,francois.vendel,etienne.delay\}@cirad.fr} \and
Institut Sénégalais de Recherches Agricoles (ISRA), CNRF, Dakar, Senegal \and
Ecole Supérieur Polytechnique (ESP), UMMISCO, Dakar, Senegal}
\maketitle              
\begin{abstract}
Cormas is an agent-based simulation platform developed in the late 90s by the Green research at CIRAD unit to support the management of natural resources and understand the interactions between natural and social dynamics.
  This platform is well-suited for a participatory simulation approach that empowers local stakeholders by including them in all modelling and knowledge-sharing steps.
  In this short paper, we present the Cormas platform and discuss its unique features and their importance for the participatory simulation approach.
  We then present the early results of our ongoing study on managing pastoral resources in the Sahel region, identify the problems faced by local stakeholders, and discuss the potential use of Cormas at the next stage of our study to collectively model and understand the effective ways of managing the shared agro-sylvo-pastoral resources.

\end{abstract}

\section{Introduction}
\label{sec:Introduction}

In recent years, there has been an increased interest in decentralization and multi-agent systems, particularly in the participatory approaches that involve stakeholders in the process of designing the model and interacting with its simulations~\cite{Becu08a,Bomm20a}.
Such techniques have been formalized in the companion modelling (ComMod) approach~\cite{Bous99a,Barr03a} that suggests to update the always-imperfect model iteratively by collecting feedback from stakeholders.
Based on the years of field expertise, the Green research unit has created the platform for multi-agent simulation Cormas (common pool resources and multi-agent simulations) which implements the commod approach and provides a unique interactive experience that is well adapted for participatory simulations.

In this short paper, we explain the ComMod approach and participatory simulations.
We briefly overview the Cormas platform and discuss its original functionalities.
We then present the early research results of our ongoing study on managing pastoral resources in Senegal and discuss the potential application of Cormas and participatory modelling at the next stage of our study.

The rest of this paper is structured as follows.
In Section~\ref{sec:Commod}, we introduce participatory modelling and the ComMod approach.
In Section~\ref{sec:Cormas}, we present the Cormas platform for agent-based simulations.
In Section~\ref{sec:Application}, we present the first results of our ongoing study on managing pastoral resources in Senegal and discuss the application of Cormas.
In Section~\ref{sec:RelatedWork}, we discuss the related work and Section~\ref{sec:Conclusion} concludes the paper.

\section{ComMod Approach and Participatory Simulation}
\label{sec:Commod}

\textit{Companion Modelling} (ComMod)~\cite{Bous99a,Barr03a} is a formalized approach that is based on a continuous feedback loop between researchers and various stakeholders.
The model should be iteratively updated based on the field situation in an endless loop \textit{``modelling''} -> \textit{``field work''} -> \textit{``modelling''} -> \textit{``field work''} -> ..., thus creating a process which produces an imperfect model but iteratively makes it ``less imperfect''.
At each iteration, the model is discussed together with stakeholders.
Their feedback and hypotheses are then used to update the model.
ComMod approach has two main objectives:

\begin{enumerate}
  \item \textit{Understanding complex environments} by making researchers interact with stakeholders and seek a mutual recognition of everyones representation of the research problems.
  \item \textit{Supporting the collective decision-making process} by incorporating differnt points of view and allowing stakeholders to learn and share knowledge with researchers but also among themselves.
\end{enumerate}

While being particularly well adapted for role-playing games, the ComMod approach has also been successfully applied with \textit{multi-agent simulations} in the form of a \textit{participatory simulation}~\cite{Becu08a,Lepa12a,Bomm14a}.
In conventional participatory modelling process, local stakeholders (the objects of study) are only contacted for data collection.
They are not involved in the other stages of the study.
ComMod is a more inclusive approach which involves stakeholders also in the process of model design, simulation analysis, and even decision making.

Collectively designing simulation models together with stakeholders is a process that could benefit from dedicated tools.
In the following section, we describe one of those tools --- a multi-agent simulation platform that was designed to implement the ComMod approach and provides several unique features to support interactive simulation. 

\section{Cormas Platform}
\label{sec:Cormas}

\begin{figure*}[htbp]
  \centering
  \includegraphics[width=.8\textwidth]{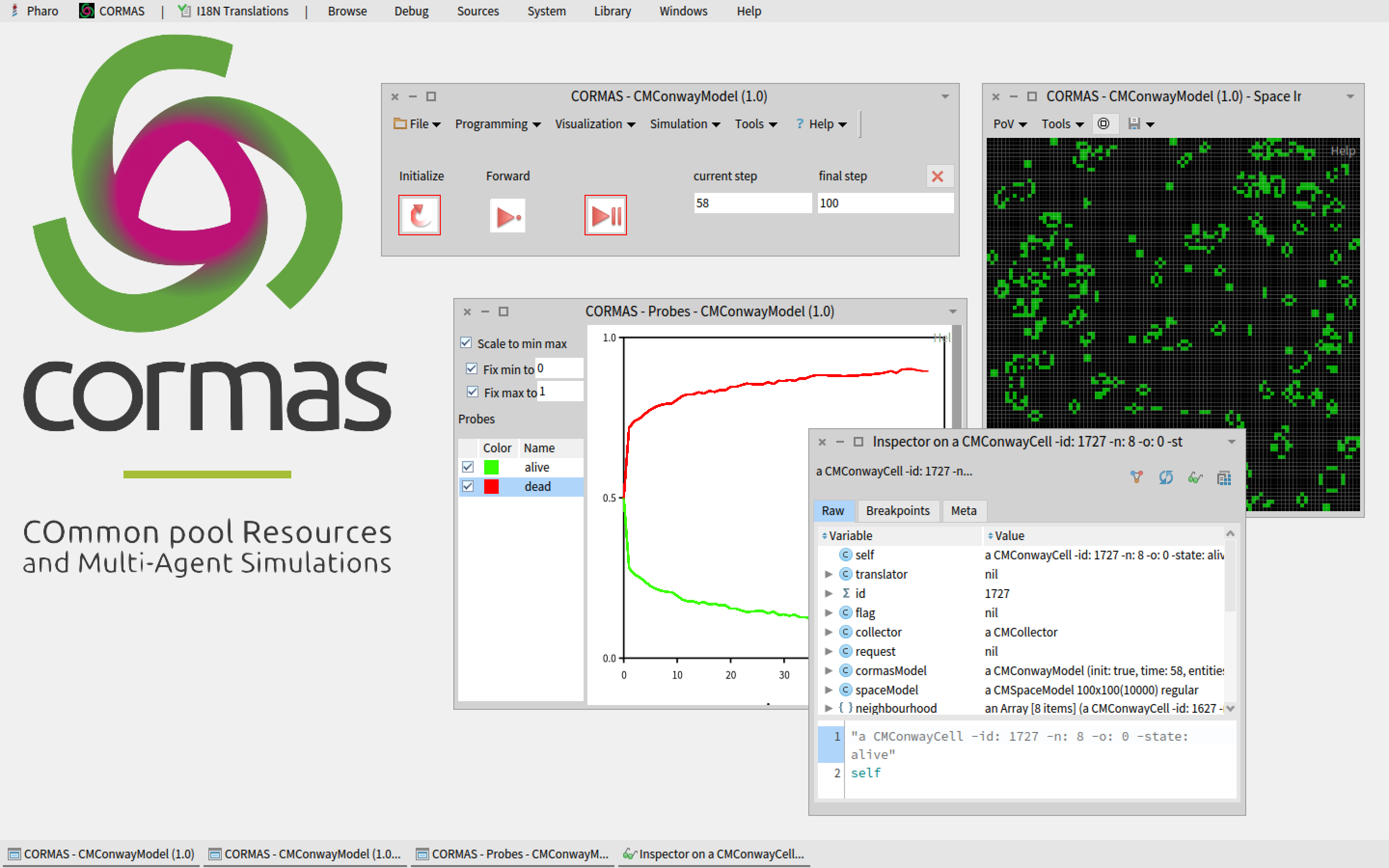}
  \caption{A screenshot of Cormas platform running the simulation of Conway's Game of Life. It contains a simulation control interface, a visualization of the space, displaying a grid cells that can be dead (black) or alive (green), a chart of two measurements (probes) - the numbers of dead and alive cells at each iteration, and the inspector window which allows user to see the properties of an individual cell and interact with it.}
  \label{fig:conway}
\end{figure*}

Cormas\footnote{\url{http://cormas.cirad.fr/}} (\textbf{CO}mmon pool \textbf{R}esources and \textbf{M}ulti-\textbf{A}gent \textbf{S}imulations) is a platform for multi-agent modelling implemented in 1997 by the Green\footnote{In 2021, the Green research unit (Gestion des ressources renouvelables et environnement) became part of the mixed research unit SENS (Savoirs, Environnement, et Sociétés): \url{https://umr-sens.fr/}} research unit of CIRAD\footnote{CIRAD (Centre de coopération internationale en recherche agronomique pour le développement) is a French agricultural research and international cooperation organization working for the sustainable development of tropical and Mediterranean regions: \url{https://www.cirad.fr/}} based on the object-oriented programming language Smalltalk~\cite{Bous98a}.
The original version of Cormas was implemented in VisualWorks Smalltalk, which is now a proprietary software system owned by Cincom\footnote{\url{https://www.cincomsmalltalk.com/main/products/visualworks/}}.
There is an ongoing effort\footnote{\url{https://github.com/cormas/cormas}} to migrate Cormas to Pharo\footnote{\url{https://pharo.org/}} --- a modern purely object-oriented programming language designed in the tradition of Smalltalk and distributed under the open-source MIT license. 

Cormas was originally designed to simplify the simulation of resource management and modelling the interactions between natural and social dynamics.
A model in Cormas can be defined either by programming it in Smalltalk (the platform provides all the building blocks that can be composed and extended) or through a graphical user interface that helps users to implement the model but the full implementation still requires them to write code.
The model can be composed of agents, which can be located in space and communicate to one another, and patches that define the space for agents. 
The platform provides various tools for analysing the model and interacting with it.
As can be seen in Figure~\ref{fig:conway}, users can visualize the space chart the progression of observed parameters, inspect specific agents or patches, and even interact with them while the simulation is runing.
Cormas also allows users to work with geographical data by importing actual maps with multiple layers of information from GIS files (geographic information systems).

\subsection{What makes Cormas unique?}

\begin{figure*}[htbp]
\centering
\includegraphics[width=.8\textwidth]{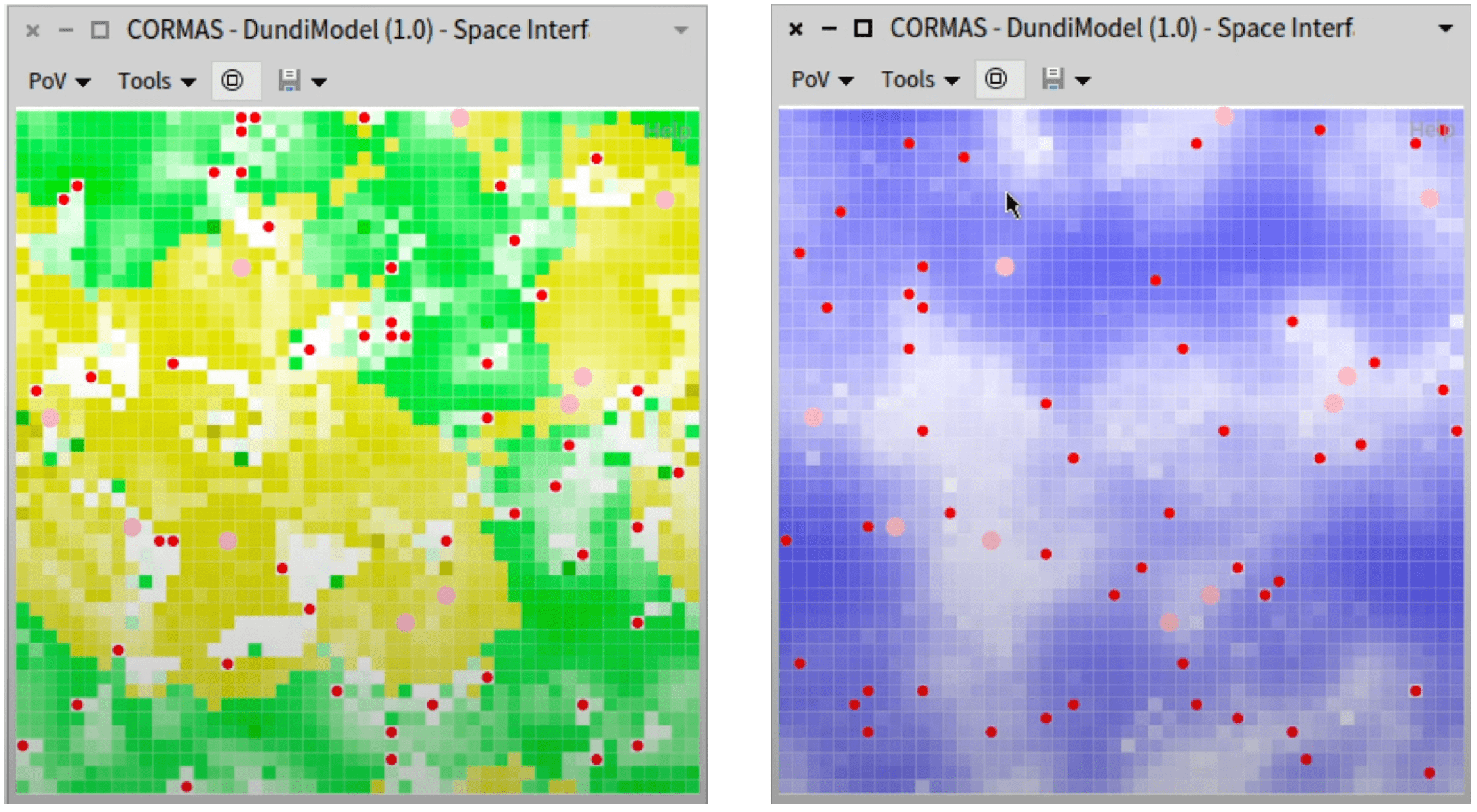}
\caption{Different "points of view" (PoV) of the same model can be displayed simultaneously. The left PoV shows the amount of grass that grows on a patch of land (intensity of green and yellow indicates the amount of fresh or dry grass) while the right PoV shows the humidity of each patch of land (intensity of blue indicates the amount of ground water). Red circles represent cows that eat grass and pink circles represent trees that consume water.}
\label{fig:pov}
\end{figure*}

Compared to other existing modelling platforms, Cormas stands out by being interactive and well adapted for the participatory modelling.
This is highlighted by its three unique features:

\begin{itemize}
  \item \textit{Different "points of view".} Cormas allows modellers to define multiple visual representation for each entity thus letting its users observe the simulation from different "points of view" at the same time (see Figure~\ref{fig:pov}).
  \item \textit{Inspecting entities.} Through the dynamic and interactive nature of Smalltalk environment, every entity (agent, patch, etc.) in a Cormas model can be inspected and controlled directly at every step of the simulation.
  \item \textit{Stepping back in time.} Cormas allows modellers to return to any past moment in the simulation runtime change the parameters or manipulate entities, and resume the simulation.
\end{itemize}

Over the years, those three features proved to be useful in the sessions of participatory simulation, allowing multiple people to observe the model from different perspectives and interact with it during the simulation runtime, thus engaging stakeholders, encouraging communication and knowledge sharing.

\section{Using Cormas to Manage Pastoral Resources in Senegal}
\label{sec:Application}

To demonstrate the application of Cormas and the companion modelling approach, in this section, we present the first insights from our ongoing study on the managemet of pastoral resources in two pastoral units: Velingara-Ferlo and Younoufere.
Both are located in the Ranerou-Ferlo department of Senegal where new reforestation initiatives are being implemented.
According to Faye~\etal~\cite{Faye01a}, the \textit{pastoral unit} is the space and the set of resources polarised by a pastoral borehole --- a hydraulic infrastructure installed to secure pastoralism and allow permanent access to water~\cite{Wane06a}.
Both units in our study are mostly populated by the Fulani people, for whom pastoralism or agro-pastoralism and mobility are characteristic of their way of life until now.
The region contains sylvo-pastoral reserves --- large protected areas intended to limit the exploitation natural resources, as well as the reforestation plots that were installed as part of the Great Green Wall Initiative\footnote{The Great Green Wall project was launched in 2007 by several African countries, aiming to create a strip of forest from Senegal to Djibouti, crossing areas mostly devoted to pastoralism: \url{https://www.unccd.int/our-work/ggwi}} in order to fight against desertification in Sahel~\cite{Dela22a}.

Both were initially under the authority of Water and Forest services. Soon after, the GGW faced to certain limits, which include replantation low rate success and social conflicts~\cite{Dela22a}.
To solve these problems, GGW accelerator was initiated in 2021 in furtherance of improving consideration to local realities and participation.

Now, there is a lot of hydraulic constructions in pastoral unit and an overlaying of power instances in natural resources management which questions relevance of actual structuration of agro-sylvo-pastoral governance.

\subsection{First Field Mission}

As part of this ongoing study, we have conducted a first field mission between mid-December 2022 and late April 2023.
We stayed 40 days in the field, engaging in participatory observation and conducting 45 semi-directed interviews by a directive sampling of local stakeholders (farmers, herders, and farmer-herders) as well as local decision makers (prefecture, Water and Forestry Services, municipality, drill community management, and the pastoral unit management).
The aim of this mission was to establish first contact with local people, learn about local practices, the transition of those practices, the management of natural resources, and local preoccupations.
The first insights of our study helped us identify the key natural resources and main problems that are faced by local communities with respect to managing those resources.
We also formulated the research questions that will guide the first iteration of companion modelling intended to better understand the identified problems and collectively find the solutions.

\subsection{First Insights}

\paragraph{Key natural resources.} Inside the Ferlo region, most local practices and especially pastoralism are driven by the availability of water~\cite{Barr83a,Diop03a}, which is the most important natural resource in our study.
It affects two secondary resources: grass and trees, which are essential to support the lives of local communities.
Grass is used as food for animals, trees are used for construction, firewood, fodder, food, and medicine~\cite{Dela22a}.

\paragraph{Problems faced by stakeholders.} By discussing with people in the local communities, we have identified the following problems that they face in both pastoral units in our study:

\begin{problems}
  \item Decline in solidarity between villages. People have reported that the relationship, trust, and communication with neighbouring villages is declining over years.
  \item Conflict between locals and transhumants (pastoral nomads) for sharing limited resources; transhumans do not respect local rules and village chiefs are losing their legitimacy.
  \item Poor collective management of shared agro-silvopastoral resources. People claim that pastoral unit is ineffective due to the lack of communication and bad management.
\end{problems}

\paragraph{Research questions.} With respect to the problems listed above, we ask the following research questions that can guide the first iteration of the companion modelling process: 

\begin{RQ}
\item What is the best scale to manage resources in the pastoral units? Should the management be delegated to the autonomous reforestation plots or is it more effective to manage it at the the level of a whole pastoral unit?
\item How to goup villages to achieve effective collaboration and resource management among them while avoiding conflicts?
\end{RQ}

We hypothesize that the informal groupments of villages, which need to be adaptive to face uncertainties (demographic, social), can be an effective form of managing agro-sylvo-pastoral resources.
Such groupment would empower local stakeholders to protect the resources that support their livelihoods, exchange knowledge, and transmit information to the higher levels of authority.
It would facilitate communication and relegitimize local and customary power.

\subsection{Next Steps with Cormas}

As part of implementing the companion modelling approach in this study, we envision two applications of Cormas for participatory simulation.

The first application is based on creating a model at the scale of a pastoral unit, where spatial patches represent different types of landscapes and are characterized by the amount and accessibility of pastoral resources.
In such a model, the agents can be individual animals or entire herds.
The interactive session based on multiple "points of view", the possibility to explore multiple scenarios would allow to collectively explore social relationships inside the pastoral unit and understand how they can be improved.
By exploring different ways to group people, we can experiment with different scales of managing resources in the pastoral unit.

The second application is based on communicating agents that represent different institutions responsible for managing the pastoral unit.
Cormas allows to represent the interactions between those agents and the sharing of information in the form of exchanging messages through the communication channels.
By collectively exploring those interactions with different groups of stakeholders, we hope to better understand the problem of poor communication and propose a feasible solution.

\section{Related Work}
\label{sec:RelatedWork}

Both the companion modelling (ComMod)~\cite{Bous99a,Barr03a} and the Cormas platform for participatory simulations~\cite{Bous98a} were introduced by the Green research unit of CIRAD.
Over the years, they found many applications in the field of natural and social sciences~\cite{Bomm16a,Bomm20a}.
LePage~\etal~\cite{Lepa12a} analysed the usage of Cormas by the research community over the period of 12 years.
They reported that the user community of Cormas was mainly interested in context-specific participatory agent-based simulation, with more than 50 training sessions organized over 6 continents. 

There were multiple applications of the participatory funtionalities of Cormas in the Sahel region.
D'aquino~\etal~\cite{Dacq03a} combined Cormas simulation together with role-playing games to empower stakeholders in the land use planning process in the Senegal River valley.
Bah~et~al.~\cite{Bah06a} used Cormas for collective mapping of land and resource usage in Thieul village, Senegal.
Their study relied on the geographical information that could be loaded into Cormas as a spatial model.
Although this study focused on different problems, it was performed in an environment similar to ours and adopted the same methodology.
The simulation of Bah~et~al. was implemented using the VisualWorks version of Cormas from 13 years ago.
We intend to replicate it using the Pharo version of Cormas.

\section{Conclusion}
\label{sec:Conclusion}

The participatory simulation approach includes stakeholders not only at stage of data collection, but also for model design, analysis, and discussion.
This encourages the sharing of knowledge between stakeholders and contributes to the collective decision making.
The Cormas platform for multi-agent simulations was designed to implement the companion modelling approach and provides some unique features for participatory simulations: multiple ``points of view'', ability to inspect and manipulate agents, and step back in time.
In our ongoing study in two pastoral units in Senegal, we adopt the companion modelling approach to improve the management of pastoral resources.
Through interaction with the local population, we have identified three problems faced by stakeholders with respect to the resource management in those pastoral units.
We have also formulated two research questions that will guide the first iteration of the future participatory simulation session designed with Cormas.
We present the vision for the two Cormas models that we plan to use at the next stage of our study and explain how the features of Cormas will contribute to the collective understanding and management of natural resources.

\section*{Acknowledgments}
\label{sec:Acknowledgments}

We are grateful to the CASSECS project for sponsoring this work and the mission of Oleksandr Zaitsev to Senegal.
We would also like to thank Dundi Ferlo project for financing the PhD of François Vendel and ISRA CNRF for hosting him in their office in Dakar, Senegal.
Finally, we would like to thank CERCIRAS COST Action (\url{https://www.cericars.org}) for paying the open-access fees for this paper and sponsoring the trip of François Vendel to present it in Cyprus.

\bibliographystyle{splncs04}
\bibliography{bibliography}

\end{document}